\documentstyle{l-aa}
\begin{document}
\input{psfig}
\def\gtsima{$\; \buildrel > \over \sim \;$}
\def\simgt{\lower.5ex\hbox{\gtsima}}
\def\capo{\\ \indent}
\def\cali{\par\noindent}
\def\bequ{\begin{center}\begin{equation}}
\def\fequ{\end{equation}\end{center}}
\thesaurus{12(12.03.1; 11.17.3)}
\title{Ionization by early quasars and Cosmic Microwave Background anisotropies}
\author{N.~Aghanim \and F. X. D\'esert \and J. L. Puget
\and R. Gispert}
\offprints{N. Aghanim }
\institute{IAS-CNRS, Universit\'e Paris XI, Batiment 121, F-91405 Orsay 
Cedex}
\date{Received date 4 April 1995 / accepted date 3 November 1995}
\maketitle
\begin{abstract}
We discuss the photoionization of the intergalactic medium by 
early formed quasars and propose a mechanism which could generate 
measurable temperature fluctuations $\delta_{T }~=~\Delta T/T$ of
 the cosmic microwave background 
(CMB).  Early quasars produce individual  ionized regions around 
themselves. We evaluate both thermal and kinetic (first-order 
Doppler) Sunyaev-Zeldovich (SZ) effects associated with those ionized
regions.  
Whereas the former is negligible, the latter induces the generation 
of detectable secondary small and medium scale CMB anisotropies.  We find 
that this effect could produce measurable individual sources 
with a $\delta_{T }\simeq 10^{-6}$ to $\simeq 10^{-4}$
on scales up to $1^{\circ}$, which introduce a non gaussian
signature in the statistics of the primordial anisotropies. The effect
of the early formed quasars is compatible with the sources detected in 
Cheng et al. (1994) MSAM CMB experiment.
\par
\end{abstract}
\section{Introduction}
In most cosmological models, some reionization of the intergalactic medium
(IGM) must have occur between recombination and the present time where most
of the gas is known to be already reionized (Gunn \& Peterson 1965). 
\par 
Several mechanisms were used to explain the ionization of the IGM. The 
possibility of photoionization has been extensively explored (Shapiro \& Giroux
1987; Donahue \& Shull 1987; Miralda-Escud\'e \& Ostriker 1990; 1992). The 
possible 
sources of photons like young galaxies, primordial stars and decaying particles
do not seem to be entirely satisfactory.
\par
Apart from photoionization, the heating of the IGM may occur in some scenarios
of galaxy formation involving explosions (Ikeuchi 1981; Vishniac \& Ostriker 
1985). 
Another possibility, which arises in hierarchical models, is ionization by high
mass stars which form in subgalactic size clumps at high redshifts (Couchman 
\& Rees 1986).
\par
A detailed discussion of the various processes of reionization would be beyond 
the 
scope of this paper. We will focus our interest on the reionization by early 
quasars and how it affects the cosmic microwave background (CMB), knowing that
Meiksin \& Madau recently (1993) have shown that the quasars might indeed 
produce enough 
ionizing photons to completely ionize the IGM.
\par
The main effect 
of the reionization is to wash out some of the temperature fluctuations of the 
CMB on small scales which undergo Thomson 
scattering on the reionized IGM (see for example Bond 
\& Efstathiou 1984; Vittorio \& Silk 1984; Tegmark et al. 1994).
\par
In the assumption of a partial reionization by quasars, we present a new
 effect 
which, on the contrary, generates secondary small scale distortions on the CMB.
\par
 At times preceding the full reionization, quasars ionize the 
gas around themselves and they become surrounded by ionized bubbles. 
 Those bubbles are moving together with the quasars in the large 
scale gravitational potential wells.  Thus they have got a bulk
peculiar 
velocity $v$, with respect to the local standard of rest in which 
the CMB is isotropic.
\par
 An effect equivalent to the Sunyaev-Zeldovich (SZ) kinetic effect 
for rich clusters creates relative temperature fluctuations which 
are proportional to the product of the Thomson optical depth 
$\tau$ by the ratio of the quasar radial velocity to the velocity of light 
$(v_{r}/c)$ -- between a line of sight crossing the 
ionized bubble and a nearby one which does not. Contrary to X-ray clusters 
, the corresponding 
thermal SZ effect, which is induced by the temperature of the 
gas is negligible in this case.
\par
By comparison, 
the Vishniac effect (1987) induces distortions due to the correlations between
the density and the velocity distributions in a {\it fully} ionized medium. The 
Vishniac effect is a second order one and it dominates on small angular scales
(a few arcminutes or a few tens of arcminutes).
\par
We investigate here a first order effect due to the bulk velocity of the 
ionized portion of the line of sight
\par
Observing the primordial anisotropies of the CMB is the only way to
constrain the various models of structure formation and give information
about the origin of the fluctuations. In fact, the 
scale-invariant spectrum of mass density is preferred in many models. The 
standard
picture of structure formation in an inflationary model naturally creates 
fluctuations with such a spectrum and a gaussian statistics. Cosmic defects, 
such
as strings, textures..., are a different source of fluctuations. 
Although they might give a scale-invariant spectrum, they introduce a 
non-gaussian characteristic to the statistics of the fluctuations.
\par
It is therefore important to determine whether or not the statistic of the 
observed primordial anisotropies is 
gaussian. In this context, it is thus also necessary to investigate mechanisms
leading to deviations from gaussian statistics in the standard model.
We find that the fluctuations induced by the early 
quasars are non-gaussian; this effect introduces a new component to the 
distortions and makes
it harder to draw conclusions about the statistical nature of the primordial
fluctuations.
\par
 Quasar formation theory is still in its infancy.  Observations 
show the existence of quasars at redshifts as high as $z~\approx~5$, although 
in hierarchical galaxy formation theories, the galaxies form 
late.  We note that a recent work (Katz et al. 1994)
shows that quasars could still be formed at high 
redshifts even in theories where galaxy formation occurs late 
such as in the hierarchical theories.  The present work  
has no strong dependence on quasar formation theory, 
the presence of any high luminosity ionizing point sources will have the same 
kind of effect on the CMB, as long as it can roughly account for the 
reionization of the universe.
\par
 For the cosmological parameters, we used $h=H_{0}/100\,{\rm km\,s^{-1}Mpc^
 {-1}}=0.5$, $\Omega_{0}=1$ (flat universe) and $\Omega_{b}h^{2}=0.0135$
 fraction of baryonic matter; 
according to the nucleosynthesis constraints we have $0.01<\Omega_{b}h^{2}
<0.015$  (Walker et al. 1991).  Hereafter, 
we assume that there are no heavy elements other than those 
created in the Big Bang (H, D, He, Li, ...). Moreover, we 
 neglect the contributions other than those induced by  
hydrogen and helium, and we assume that the quasars emit 
their radiation isotropically, producing spherical ionized bubbles. Possible 
geometrical effects were neglected in the present work. 
 Owing to the high redshift range of interest, the hypothesis 
of homogeneity for the IGM gas is valid.
\par
These approximations are justified in view of other uncertainties on the
reionization process. The goal of this paper is to show that a non
negligible effect is induced by this process and not to compute a
detailed model.
\par
In section 2 we discuss the relevant properties of an isolated 
bubble around a quasar and we compute the distortion 
of the CMB induced by such a bubble. In section 3 we compute 
the statistics over the whole sky of the distortions induced. 
In section 4 we present 
the results and discuss them in relation to recent CMB measurements.
\section{ Isolated early quasars and their effect on the CMB}
The specific quasar model we use is a good illustration for effects induced 
by other possible sources of ionization like young ultraluminous galaxies for
example.
\par
We define the period of {\it isolated quasars} as  
the redshift range where most of the baryonic matter is still neutral
while early quasars appear.
\par
 An isolated quasar emits strong radiation -- in particular in 
the UV range -- that can ionize the neutral intergalactic gas around 
 it at great distance, thus creating an ionized sphere. 
\subsection{ Ionizing radiation production rate}
We first compute the size of the ionized sphere surrounding the 
quasar.  Knowing the quasar specific luminosity in the $B$ band, 
we deduce the 
emission rate $S$ of UV photons ionizing the hydrogen atoms of the IGM.
\par
 The intrinsic spectrum of individual quasars in the ionizing 
UV domain is still uncertain but a good approximation is to use 
a power law extrapolation of the optical flux.
\par
 Several observations have been made in different spectral ranges (optical, 
 X-rays ...) and constraint
the values of the average spectral indexes in each range. A review of these
observations was reported for 
example by Bechtold et al. (1987) who considered three spectral forms.
We adopt the model consistent with the ``medium'' spectrum 
(Bechtold et al. 1987) which gives for the continuum spectral energy 
distribution radiated by quasars:
\bequ
L_{\alpha}~\propto~\nu^{-1.5}~(\lambda<1216\,\mbox{\AA}).
\fequ
\par
The luminosity function for quasars defined in Sect. 3.1 is 
written as a function of $L_{\alpha}$ the luminosity per spectral 
interval at the Lyman $\alpha$ frequency.  It is therefore necessary to 
write $S$ the UV emission rate -- expressed in number of ionizing 
photons per unit time -- as a function of $L_{\alpha}$
\bequ
S~=~\int_{\nu_{ion}}^{\infty}\frac{L_{\alpha}}{h_{pl}\nu}~d\nu,
\fequ
where $h_{pl}$ is the Planck constant and $\nu_{ion}$ is the ionizing
frequency for hydrogen.
\par
Since the integral is dominated by the lower limit in frequency (ionizing 
limit for the hydrogen), the ionizing flux is not critically dependent
on the assumed spectral shape.  We get $S~=~AL_{\alpha}$    
where $A$ is a constant depending on the normalization of the spectrum. 
We introduce the following notation:
a characteristic luminosity $L^{*}_{\alpha}(z_{c})~=~L^{*}_{c}$
which is independent on the quasar age and will be introduced 
in Sect. 3.1, and defining the ratio $l_{Q}~=~L_{\alpha}/L^{*}_{c}$, 
one gets 
\begin{equation}
S~=~\frac{L^{*}_{c}}{\alpha h_{pl}}\left(\frac{\lambda_{Ly}}
{\lambda_{ion}}\right)^{-\alpha}l_{Q}.
\end{equation}
$\alpha=1.5$ is the spectral index of the quasar spectrum in the range of 
interest.
\par
\subsection{Temperature of the bubble}
 If the intergalactic matter around the quasar -- essentially composed 
of hydrogen and helium -- is ionized by the emitted radiation, the ionization of 
a hydrogen atom produces a photoelectron with an average kinetic 
energy defined by $\overline{E_{c}}= \overline{E_{UV}}-13.6\,\mbox{eV}$, 
where $\overline{E_{UV}}$ is the average energy of the ionizing photon.
\par
 Since electrons and protons rapidly thermalize owing to the 
short time scale of the Coulomb interactions, the electrons transfer 
about half of their energy to the protons, and the temperature 
of the plasma is thus given by $T=\overline{E_{c}}/3k_{b}$ as long as 
the cooling by recombination is negligible; here $k_{b}$ is the Boltzmann
constant.
\par
 We can compute the average energy of the ionizing photon
$\overline{E_{UV}}=h_{pl}\overline{\nu}$
  knowing the quasar spectrum and assuming that every photon 
is absorbed sooner or later:
\bequ
\overline{\nu}~=~\frac{\int_{\nu_{ion}}^{+\infty}F_{\nu}d\nu}{
   \int_{\nu_{ion}}^{+\infty}\nu^{-1}F_{\nu}d\nu}.
\fequ
 We find $\overline{E_{UV}}\approx 40\,\mbox{eV}$, which leads to 
 $T\approx 10^{5}\,\mbox{K}$ 
for the plasma temperature. A photoelectron even if it has an energy 
larger than $13.6\,\mbox{eV}$ will have a very low probability to make a second
ionization, the propagation of the ionization front being much faster than 
the electron velocity. Thus the number of ionization is very nearly equal 
to the number of ionizing photons. In that context, taking into account
the presence of helium is equivalent to a $10\%$ increase of the baryon 
density with hydrogen only. Therefore, in the following, we use $\Omega_
b h^2=0.015$. 
\par  
In the bubbles surrounding the quasars, the gas is fully ionized, therefore
there is no cooling due to neutral atoms. 
The ionized gas will cool by 
bremsstrahlung (free-free radiation), but, the cooling time is 
longer than the age of the universe. 
\par
Likewise, we investigate the effect of the inverse Compton cooling and
find that it will become important for redshifts greater than
$10\, h^{2/5}$ (cooling time equal to the age of the universe), or for 
$h=0.5$, $z>7.6$.
\par
We check that the temperature of the bubble decreases by a factor 2 at
$z=10$, the Compton cooling does not change drastically the temperature.
Since the exact temperature is not critical, we assume that it remains 
constant at $T\approx 10^{5}\,\mbox{K}$ but undergoes the adiabatic cooling 
due to expansion.
\par
This approximation breaks down for redshifts significantly larger than 10.
\par
\subsection{Radius and angular size}
We now compute the radius of the ionized region induced by the 
quasar and first the Str\"{o}mgren sphere (stationary ionization 
bounded H{\sc ii} region).
\par
The proper radius $R_{s}$  of the Str\"{o}mgren sphere is derived from 
\bequ
\frac{4\pi}{3}R_{s}^{3}\alpha_{r}n_{e}^{2}=S,
\fequ
where $n_{e}$ is the electronic density $n_{e}=n_{0}( 1 + z )^{3}$ 
($n_{0}=1.13\,10^{-5}\,\Omega_{b}h^{2}\,{\rm cm^{-3}}$ is the present 
electronic density). 
 This approximation is valid as long as the photoelectron energy 
does not exceed the photoionization energy that is to 
say that a photoelectron is unable to produce additional ionization.
The recombination coefficient $\alpha_{r}$ to all hydrogen levels is given 
by (Hummer \& Seaton 1963)
\[\alpha_{r}~=~1.627\,10^{-13}\,T_{4}^{-1/2}\,\times \]
\bequ
(1- 1.657\,\log T_{4}+ 
0.584\,T_{4}^{1/3})\,{\rm cm^{-3}\,s^{-1}},
\fequ
where $T_{4}$ is the temperature of the ionized IGM inside 
the Str\"{o}mgren sphere in units of $10^{4}\,\mbox{K}$; thus:
$\alpha_{r}\approx 2\,10^{-14}\,{\rm cm^{3}\,s^{-1}}$.
\par 
Finally, the Str\"{o}mgren radius is given by:
\bequ
R_{s}~=~\left(\frac{3S}{4\pi\alpha_{r}n_{0}^{2}}\right)^{1/3}
(1+z)^{-2}.
\fequ
On the other hand, 
let us assume that a quasar turns on at a redshift $z_{on}$ and 
turns off at $z_{off}$, in the assumption of a lifetime 
$t_{q}$ short compared to the expansion time, we have 
$1 + z_{on}\approx 1 + z_{off}$.  The proper 
radius $R$ of an ionized bubble 
that is produced around it at a given time $t$ (when it has not 
reached the stationary Str\"{o}mgren sphere state) is computed 
by writing that every UV photon emitted by the quasar is absorbed 
through the photoionization of a neutral H atom; therefore the 
main quantity that rules the size of the bubble is the total 
number of ionizing photons that are emitted by a quasar during 
its lifetime: this is given by the product $St_{q}$.
\par
The proper radius $R$ will reach a maximum value at the redshift 
$z_{off}$ (turn off of the quasar) $R_{max}(L_{\alpha},z_{off}$); and 
then it undergoes the average Hubble expansion (the dynamical expansion
due to the internal pressure of the bubble can be neglected).
\par
This maximum radius is the solution of:
\begin{equation}
\frac{4\pi}{3}R_{max}^{3}n_{0}(1+z_{off})^{3}~=~St_{q}.
\end{equation}
At a given redshift $z$ smaller than $z_{off}$, the expansion leads to the 
radius given by the following relation:
\bequ
R(L_{\alpha},z)~=~\left(\frac{3St_{q}}{(4\pi)n_{0}}
\right)^{1/3}(1+z)^{-1},
\fequ
defining $R_{c}=(3AL_{c}^{*}t_{q}/4 \pi n_{0})^{1/3}$
a characteristic radius associated with the characteristic luminosity, 
$L_{c}^{*}$ and the quasar lifetime $t_q$ we finally have:
\begin{equation}
R(l_{Q},z)~=~R_{c}l_{Q}^{1/3}(1+z)^{-1}.  
\end{equation}
For  $t_{q}= 10^{8}\,\mbox{yrs}$, we find for our choice of cosmological
parameters $R_{c}=10\,\left (\frac{\Omega_{b}h^{2}}{0.015}\right)^
{-1/3}\,\left(\frac{\Omega_0}{1}\right)^{-1/6}\,\mbox{Mpc}$.
\par   
The comparison of the two computed radii (Eq. (7) and (10)) shows that for 
redshifts less than $z_{eq}$ where $1+z_{eq}=(t_{q}\alpha_
{r}n_{0})^{-1/3}$, ($\approx 47$ for $\Omega_{b}h^{2}=0.0135$ 
and $t_{q}= 10^{8}\,\mbox{yrs}$) the ionized bubble is smaller than 
its Str\"{o}mgren radius and therefore recombinations can be neglected.
\par
The angular size (radius) of the ionized bubble is $\theta=R/D_{A}$, 
where $D_{A}$ is the angular distance; for a flat 
universe it is given by
\bequ
D_{A}~=~\frac{2c}{H_{0}}(1+z)^{-2}[1+z-\sqrt{1+z}].
\fequ
Finally, by defining $\theta_{c}=R_{c} H_{0}/ 2c$, which is given by 
$\theta_c\approx 3
\left(\frac{\Omega_{b}h^{2}}{0.015}\right)^{-1/3}\left(\frac{\Omega_0}
{1}\right)^{-1/6}\left(\frac{h}{0.5}\right)\,\mbox{arcmin}$, 
and using Eq. (10) we get
\begin{equation}
\theta~=~\theta_{c} l_{Q}^{1/3} [1-(1+z)^{-1/2}]^{-1},
\end{equation}
which is proportional to $(\Omega_b h^2)^{-1/3}\,h$, in a flat universe.
Figure 1 shows the angular radius of an ionized bubble as a function 
of redshift for different values of the quasar luminosity and lifetime.
\par
\begin{figure}
\psfig{file=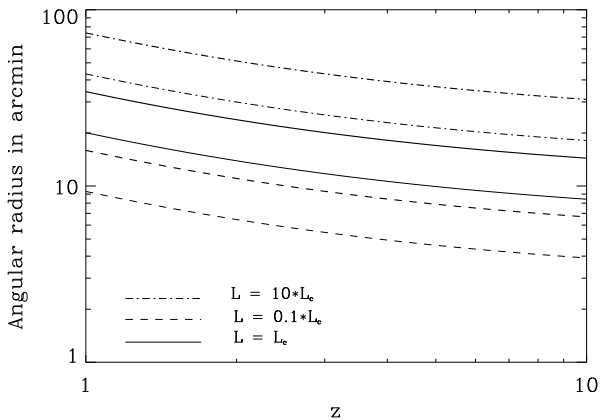,width=\columnwidth}
\caption[ ]{Angular radius of an ionized bubble as a function of the 
redshift for three values of the quasar luminosity, where $L_{c}^{*}$ 
is a characteristic 
luminosity defined Sect. 3.1. The angular radius varies as $L^{1/3}$ see Eq. 
(12). The series of thin lines stands for the case of the angular 
radius computed with a quasar lifetime $t_{q}= 10^{7}\,\mbox{yrs}$, whereas, 
the 
thick lines are for $t_{q}= 5\, 10^{7}\,\mbox{yrs}$. The angular radius 
increases with $t_q^{1/3}$ (Eq. (12)).}
\end{figure}
\cali
\subsection{Optical depth}
Knowing the proper radius of the ionized sphere, we can compute 
the Thomson optical depth for the line of sight going through the
center of the bubble: 
\bequ
\tau_{m}~=~2\int_{0}^{R}\sigma_{T}n_{e}dl,
\fequ 
$\sigma_{T}= 6.65\,10^{-25}\,{\rm cm^{2}}$ is the Thomson cross section.  
We have thus:
\bequ
\tau_{m}~=~2 \sigma_{T}n_{e}(z)R(L_{\alpha},z),
\fequ 
putting $\tau_{c}=\sigma_{T}n_{0}(z)R_{c}$ the characteristic optical
depth obtained for $L_{c}^{*}$, we have 
\begin{equation}
\tau_{m}~=~\tau_{c}l_{Q}^{1/3}(1+z)^{2}.
\end{equation}
The average optical depth over the spherical structure is $\tau_{a}~=~
\sigma_{T}n_{e}(z)\overline{R(L_{\alpha},z)}$ with $\overline{R(L_{\alpha},z)}
=\frac{4}{3}R(L_{\alpha},z)$, therefore
\begin{equation}
\tau_{a}~=~\frac{2}{3}\tau_{m},
\end{equation}
and it scales like $(\Omega_{b}h^{2})^{2/3}\,\Omega_0^{-1/6}$.
Figure 2 shows the dependence of the bubble optical depth with both the 
lifetime of the quasar and its luminosity, as a function of the redshift.
One should notice in Fig. 2 that the typical optical depth of $10^{-3}$
(comparable to those of rich clusters) are reached for moderate redshifts.
\par
\begin{figure}
\psfig{file=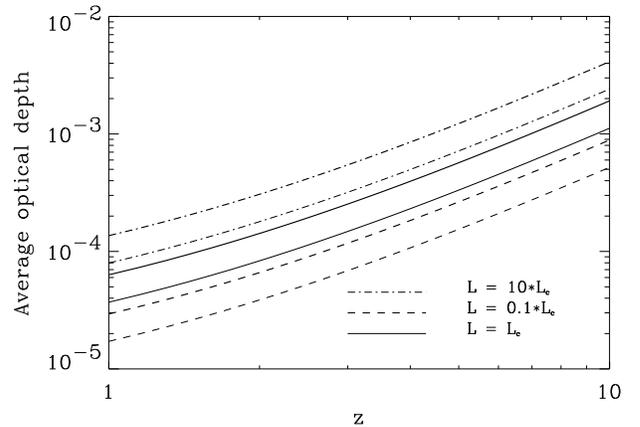,width=\columnwidth}
\caption[ ]{Average optical depth of an ionized bubble versus redshift for 
different values of the luminosity with lifetime $t_{q}=10^{7}\,\mbox{yrs}$ 
in 
the case of the thin lines. The thick lines represent the increase due to the 
increase of the lifetime $t_{q}=5\, 10^{7}\,\mbox{yrs}$.}  
\end{figure}
\subsection{ Thermal Sunyaev-Zeldovich effect} 
Compton scattering of the CMB radiation on hot free electrons 
imprints on the incident photon spectrum a signature known as 
the Sunyaev-Zeldovich thermal effect (SZ) as proposed by Zeldovich 
\& Sunyaev (1969) and Sunyaev \& Zeldovich (1972; 1980) by redistributing 
the photons in the spectrum with a general shift toward higher 
frequencies. This is an effect that has initially been proposed 
for clusters of galaxies, and it has already been observed in the direction 
of several rich clusters (see for ex Birkinshaw 1990; Birkinshaw et al. 
1991; Wilbanks et al. 1994).
\par
The comptonization parameter, 
$y$, which characterizes the SZ distortion, depends on the bubble's 
characteristics  and is essentially 
proportional to the line integral of the electron pressure along the
line of sight. For an isothermal sphere, it is given by:
\bequ
y~=~\frac{k_{b}T_{e}}{m_{e}c^{2}}\tau;
\fequ
$T_{e}$ is the electronic temperature and $m_{e}c^{2}$ is the electron rest 
mass.
In the Rayleigh Jeans domain the distortion amounts to $\frac{\Delta T}{T}=-
2y$.
It will be shown to be negligible because of the rather low temperature of 
the plasma in the bubble.
\par
\subsection{ Kinetic Sunyaev-Zeldovich effect} 
Another type of interaction between the bubble and the CMB photons 
is the SZ kinetic effect (first-order Doppler effect).
\par
 The natural reference frame to describe the SZ effect is the local 
standard of rest defined by the very large scale distribution of matter,
i.e the frame in which the CMB is isotropic.
 If the bubble is at rest in this frame, then the thermal SZ 
effect (second order effect with respect to the average velocity of the 
electrons) is the only signature imprinted on the spectrum.
\par
 The significance of the so called kinetic effect has been given 
by Sunyaev \& Zeldovich (1972); the motion of a cluster with 
respect to the background leads (due to the Doppler effect) to 
an additional change of the radiation temperature 
in its direction, because of the finite optical depth associated 
to the bubble.  The spectrum of the distortion is indistinguishable 
from a primordial anisotropy spectrum which is equivalent to a simple
temperature change. The observed radiation 
temperature is changed by $\delta_{T}=\Delta T/T = -(v_{r}/c)\tau$, in a 
direction crossing the bubble where $\tau$ is the Thomson optical depth  
and $v_{r}$ is the radial component of the bubble peculiar velocity 
(a positive one corresponds to a recession velocity). 
\par
 In order to evaluate the relative temperature fluctuations due 
to the SZ kinetic effect associated with the spheres of plasma 
around early quasars or left by them after they stop emitting ionizing
radiation, we need the optical depth (see Sect. 2.4) and the radial 
velocity of the bubbles.
The average anisotropy associated with an ionized bubble is given by:
\begin{equation}
\delta_{T}~=~\frac{2}{3}\frac{v_{r}}{c}\tau_{c}
l_{Q}^{1/3}(1+z)^{2};
\end{equation}
this gives us a direct relation between the luminosity of a quasar, 
its radial velocity, the redshift and $\delta_{T}$.  As regards the 
cosmological
parameters, we have $\delta_T \propto (\Omega_b h^2)^{2/3}\,\Omega_0^{-1/6}$.
\par
\subsection{  Velocities}
In standard cosmological models, large scale velocities are assumed 
to be induced by gravity from mass density fluctuations: infall 
into over dense regions and outfall out of underdense regions. 
 According to the linear gravitational instability theory, the 
rotational part of the velocity field is diluted by the expansion 
so that the present large scale cosmic velocity field is expected to be a 
potential flow.  The velocity dispersion of the large scale structures 
is computed in the assumption of the linear theory; it is given by 
$\sigma_v(z)=\sigma_{0v}
(1+z)^{-1/2}$ (Peebles 1980; 1993), where $\sigma_{0v}=\sigma_v(z=0)$.
\par
 In the following, we will assume that the bubbles produced by quasars 
 are random substructures 
in the linear potential of large scale structures and thus in motion with them.
The distribution 
of their radial velocities is therefore given by the large scale velocity 
field distribution.  The velocities are dominated by the maximum 
of $\sigma_{0v}(L)$, where $L$ is the corresponding scale, 
the maximum takes place for scales about $10\,\mbox{Mpc}$ which is still 
linear at redshifts of interest $(z > 5)$.  The velocity distribution 
is directly related to the distribution of primordial density 
fluctuations during the linear stage; in the assumption of  
gaussian statistic for the fluctuations, the peculiar one dimensional 
velocity distribution of the quasars and thus the ionized bubbles 
is also gaussian, with a velocity dispersion $\sigma_v(z)$.
\par
Numerical simulations have been done to evaluate $\sigma_{0v}$ and showed
 for different cosmological models 
that it was in a range about $700\,\mbox{km/s}$ to $200\,\mbox{km/s}$ 
depending 
on the model parameters (Rhee 1993; Croft \& Efstathiou 1994). 
\par
The observed bulk flow within a sphere out to 6000 km/s is 350 to 400 km/s
(Faber et al. 1993). We take 
here a semi empirical approach using the $z$ dependence given by the linear 
theory and a conservative value of  $\sigma_{0v}=300\, \mbox{km/s}$. The 
coherence length of the 
bulk flow could be significantly larger than in the Cold Dark Matter (CDM)
 model and thus lead
to a larger effect. 
\par

The ratio of SZ thermal effect to kinetic effect is independent on the optical
depth and given by:
\bequ
\frac{k_{b}T_e}{m_{e}c^{2}}\frac{c}{v_{r}}~=~1.7\,10^{-2}\left(
\frac{T_{e}}{10^5\,\mbox{K}}\right)\left(\frac{v_{r}}{300\,\mbox{km/s}}
\right)^{-1}.
\fequ
Hence the thermal effect is 
negligible compared with the kinetic effect.  This is opposite to what
happens for clusters, because the intracluster gas is much hotter 
($\approx10^{8}\,\mbox{K}$).
\par
Figure 3 shows the temperature fluctuation taking for the value of the 
velocity, the velocity dispersion,
as a function of redshift for different luminosities and lifetimes.
\par
\begin{figure}
\psfig{file=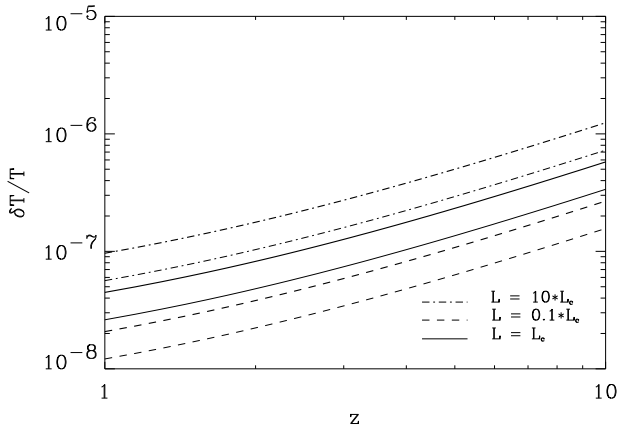,width=\columnwidth}
\caption[ ]{Temperature fluctuation associated with an ionized bubble 
versus the redshift for a radial velocity 
given by the velocity dispersion (Sect. 2.7) and 
different quasar luminosities. Thin lines are obtained with quasar lifetime 
$t_{q}=10^{7}\,\mbox{yrs}$ and thick ones with $t_{q}=5\, 10^{7}\,\mbox{yrs}$.}
\end{figure}
\section{ Quasars distribution}
\subsection{ Quasars luminosity function}
The statistics of the measurable bubbles can be deduced if one knows the 
distribution and evolution of the ionizing sources. This is done when we
take into account the quasars distribution given by 
their luminosity function $\Phi(L_{\alpha},z)$ which expresses 
for a given redshift $z$ the comoving space density 
of quasars as a function of the luminosity.
\par
 The quasars luminosity function is not well determined yet, 
in particular at high redshifts ($z > 3$), owing to the small 
number of observed sources. Therefore, we will consider different 
expressions for the luminosity evolution depending on the redshifts 
range of interest.
\par
According to recent surveys (Boyle, Shanks, \& Peterson 1988; Boyle, Jones 
\& Shanks 1991), a pure luminosity evolution model -- in
which the comoving number density of quasars is assumed to be statistically
constant with respect to the time while the luminosity varies with $z$ --
describes well the properties of the quasars population up to $z\approx 3$.
For $z > 3$, following Warren et al. (1991) and Meiksin \& Madau (1993) 
we introduce a pure density
evolution model to describe the population beyond $z\approx 3$ which 
states that the comoving number density decays exponentially by a factor 
of $2$ per unit redshift (see Hartwick \& Shade 1990 for discussion of the
quasars luminosity function).
\par
The luminosity function can be written as: 
\bequ
\Phi(L_{\alpha},z)~=~\phi(L_{\alpha},z)f(z,\tilde{z}),
\fequ
where
\bequ
\phi(L_{\alpha},z)~=~\frac{\Phi^*}{L_{\alpha}^*(z)}\left\{\left[
\frac{L_{\alpha}}{L_{\alpha}^{*}(z)}\right]^{\beta_{1}}~+~\left[
\frac{L_{\alpha}}{L_{\alpha}^{*}(z)}
\right]^{\beta_{2}}\right\}^{-1}
\fequ
is the best fit to the quasars luminosity function up to $z \approx 2.9$,
(Boyle 1991), and $f$ is a luminosity evolution function 
(see below).
\par
In a cosmological model with $\Omega_{0}=1$ and $H_{0}=50\,{\rm km\,s^{-1}\,
Mpc^{-1}}$, the "best-fit" parameters for the 
model are (Boyle 1991) $\Phi^*= 6.5\,10^{-7}\,{\rm Mpc^{-3}}$, $\beta_1=
3.9$, $\beta_{2}=1.5$.
\par
The absolute $B$ magnitude, $M_{B}$, variation with redshift is expressed 
by $M_{B}(z)=M_{B}^{*}(z=0)-2.5 k_{L}\log(1+z)$, where the $B$ magnitude at 
the present epoch is $M_{B}^{*}(z=0)=-22.4$ and $k_{L}=3.45$ ("best-fit"
parameter). This defines a characteristic 
luminosity $L_{\alpha}^{*}(z)=10^{-0.4 M_{B}(z)+20.29}\,{\rm erg\,s^{-1}\,Hz^
{-1}}$.
\par
It finally  gives 
\bequ
L_{\alpha}^{*}(z)~=~\left\{\begin{array}{ll}L_{\alpha}(0)(1+z)^{k_{L}} & 
\mbox{for $z < z_c$} \\ 
L_{\alpha}^{*}(z_{c})=L^*_c & \mbox{otherwise,}\end{array}
\right.
\fequ
where $z_{c}$ is a characteristic redshift, $z_{c}=1.9$ ("best-fit" parameter).
The luminosity evolution models are expressed by the function 
$f(z,\tilde{z})$
depending on the value of characteristic redshift $\tilde{z}$. In the constant 
comoving range for $z < \tilde{z}= 3$ we have $f(z,\tilde{z})=1$, while in the 
exponential decay mode, for $z > \tilde{z}= 3$, $f(z,\tilde{z})=
exp(-\mu(z-\tilde{z}))$; $\mu$ is the decay constant. 
The factor $2$ decline on the 
number of the observed quasars a high redshifts gives $\mu = 0.69$.
\par
The comoving spatial density of the quasars in function of the 
redshift in the range of interest is therefore given by the following 
expression:
\[ \phi(L_{\alpha},z)~=~\frac{\Phi^*}{L_c^{*}}exp(-\mu(z-3)) \,
\times \]
\bequ
\left\{\left[\frac{L_{\alpha}}
{L_c^{*}}\right]^{\beta_{1}}~+~\left[\frac{L_{\alpha}}
{L_c^{*}}\right]^{\beta_{2}}\right\}^{-1} \: if\, z>\tilde{z}=3.
\fequ
\subsection{ Quasars formation rate}
The luminosity function gives the number of the observed quasars 
but it can be associated with very different quasars production rate
depending on their lifetimes. This is important because we need to 
include ionized bubbles created by already {\it dead} quasars.
The number of bubbles at a given 
time is thus larger than the number of radiating sources at the same
moment. Only under the assumption
 that quasars have a lifetime much longer than the age of the
universe, does the luminosity function 
represent the number of ionized spheres.  The size of the bubble 
will increase faster than $(1+z)^{-1}$ if the quasar lifetime is infinite.
\par
A more likely assumption is that the quasar lifetime is 
short relative to the age of the universe. In this case, although the 
quasar does not emit any more ionizing photons, it has already produced 
a long lived bubble that undergoes expansion as we have seen in Sect. 2.3. 
\par
 We define the rate of production of quasars $Q(L_{\alpha}, z)$ as
\begin{equation}
{\it Q}(L_{\alpha},z)~=~\frac{\Phi(L_{\alpha},z)}{t_{q}},
\end{equation}
where we assume that the luminosity function is constant during 
the short quasar lifetime 
$t_{q}$. This quantity will thus express the spatial 
density of ionized bubbles depending on the redshift and the 
luminosity of the quasars that have formed them.
\par
 The theory of quasar formation and evolution is still very 
uncertain, therefore the quasar lifetime $t_{q}$ was alternatively given 
two values 
$t_{q}=10^{7}\,\mbox{yrs}$ and $t_{q}=10^{8}\,\mbox{yrs}$ (Padovani et al. 1990).  
We note that decreasing the lifetime, 
leads to the production of smaller and more numerous bubbles.
\par
\subsection{ Porosity parameter}
The Gunn-Peterson (1965) test for hydrogen shows that the universe is ionized 
at redshift greater than 5. Recent measurements of the Gunn-Peterson effect
for helium based upon the Hubble Space Telescope FOC observations by
Jakobsen et al. (1994), indicate that the He{\sc ii} reionization of the 
universe is
partial until $z \approx 3$ (Madau \& Meiksin 1994).
We suppose that the photoionization of the 
intergalactic medium is due to the production of H{\sc ii}
regions associated with UV sources (quasars); these 
regions expand and must overlap by $z_{ion}\approx 5$.

 In order to evaluate the redshift $z_{ion}$ of the reionization 
of the universe for 
hydrogen, one can compute the porosity parameter $
{\it P}$.  It is the product of the volume associated with an individual 
ionized bubble by the spatial number density of the bubbles
$$
{\it P}~=~\int_{z}^{+\infty}\frac{dt}{dz_{off}}dz_{off}
$$
\begin{equation}
\int\frac{4\pi}{3}R^{3}(L_{\alpha},z)
\phi(L_{\alpha},z_{off})\,dL_{\alpha}.
\end{equation}
 Since every UV photon must be absorbed in the IGM, the volume 
filling factor of ionized hydrogen is equal to the porosity parameter 
${\it P}$ (and not to $1-\exp(-P)$, which would be the case if the physics
would fix the radius of the bubble rather than the ionized volume).  Thus 
the complete reionization of the IGM occurs for 
${\it P }$ reaching 1. Figure 4, which is plotted using the set of parameters
given in Sect. 1,
shows that the assumptions from previous 
sections lead to $z_{ion}\approx 5.7$. Furthermore, the choice of the 
cosmological parameters influence the value of the porosity parameter ($P
\propto
(\Omega_bh^2)^{-1}\,h^{-1}\,\Omega_0^{-5/6}$). The reionization 
redshift is independent on the quasar lifetime as long as the quasars 
production rate and associated lifetime leads to the observed luminosity 
function. Our assumptions are therefore compatible with the observed
reionization of the universe. 
\par
The integration over the luminosity takes into account the total 
contribution of the quasars to the reionization.
\par
At redshifts where ${\it P }\ll1$, the universe is only partially ionized 
and the produced ionized bubbles are isolated in a neutral universe. 
All the quantities defined Sect. 2 (angular size, optical 
depth and temperature fluctuation) can thus be computed for these 
bubbles.  Given the luminosity function and the rate of production 
of quasars we are able to give statistical properties for the 
relevant quantities; we compute the number of bubbles as a function of
the redshift, their angular size and the CMB temperature 
fluctuation they induce, the surface 
covering factor...
\par
The induced CMB temperature fluctuations will be dominated by a redshift 
range in which individual ionized bubbles induce a significant 
Doppler effect, this is when the hydrogen is still mostly neutral that is
for redshifts $z>z_{ion}$.
\par
\begin{figure}
\psfig{file=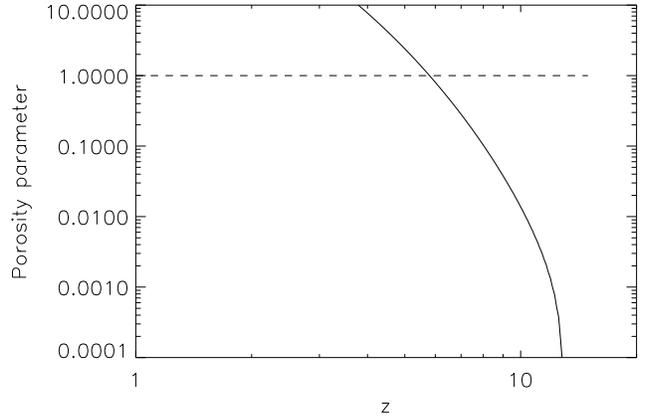,width=\columnwidth}
\caption[ ]{The porosity parameter is plotted ({\it solid curve}) as a function 
of the redshift. The horizontal {\it dashed line} stands for the reionization.
Porosity parameter is independent on the lifetime of the quasars.}
\end{figure}
\subsection{  Number of bubbles}
Starting from the quasars production rate, 
we first compute the number of ionized bubbles $N$ as a function of 
their associated temperature fluctuations $\delta_{T}=\Delta T/T$, 
their angular sizes $\theta$, redshift 
$z$, per units of solid angle.  Knowing the number density of bubbles 
per units of volume, $N$, which is given by:
\bequ
N~=~\int_{0}^{t}{\it Q}(L_{\alpha},z)dt\,.
\fequ
We derive the number of bubbles $dN=NdV$ in the elementary volume 
$dV=a^{3}r^{2}\,drd\Omega$ with $adr=cdt$ and $a$ the scaling factor 
$a=(1+z)^{-1}$, $d\Omega$ is the viewing solid angle; with 
$r=3ct_{0}(1-(1+z)^{-1/2})$ assuming an Euclidean universe.
\[ \frac{d^2 N}{dzd\Omega}(L_{\alpha},z)~=~6\left(\frac{c}{H_0}\right)^{3}
\frac{t_{0}}{t_{q}}k(z) \, \times \]
\bequ
\int_{z}^{+\infty}\Phi(L_{\alpha},z_{off})
(1+z_{off})^{-5/2}dz_{off},
\fequ
here
$k(z)=(1+z)^{-9/2}[1+(1+z)^{-1/2}]^{2}$.
\par
Using the quantities computed in Sect. 2 (radius, angular size 
and optical depth of the ionized bubbles around quasars) which are function 
of the redshift and the quasar luminosity, we can compute the number 
of bubbles at redshift $z$ with angular size $\theta$ and absolute value
of the temperature 
fluctuation $|\delta_{T}|$ in the universe, per unit of solid angle 
as follow:
$$
\frac{d^{4}N}{d\Omega\,dz\,d\ln\theta\,d\ln\ |\delta_T|}~=~2\, \times\,6
\left(\frac{c}{H_0}\right)^{3}\frac{t_{0}}{t_{q}}f(z)\,\theta\,|\delta_{T}|\,
\times
$$
\[
\int\int\phi(l_{Q})g(v_{r},z)\,{\bf \delta}[l_{Q}-l_{Q}(\theta,z)]\,dl_{Q}
\]
\bequ
{\bf \delta}[\delta_T -\delta_T (v_r,l_Q,z)]dv_{r}\,\int_{z}^{+\infty}
p(z_{off})\,dz_{off};
\fequ
where ${\bf \delta}$ is the Dirac function, $g(v_r,z)$ is the gaussian 
distribution for the radial peculiar velocities of the quasars,
\par
$f(z)=(1+z)^{-9/2}[1+(1+z)^{-1/2}]$,
\par
$p(z_{off})=(1+z_{off})^{1/2}\exp[-0.69(z_{off}-3)]$
\par 
and $l_{Q}(\theta,z)$ is derived from Eq. (12) and $v_r$ from Eq. (18).
\par
There are as many negative sources as positive ones, therefore the factor 2 is
introduced when we take into account the negative and positive radial 
velocities of the 
bubbles and give the counts as a function of $|\delta_{T}|$. We note
that in a flat universe, the number of bubbles scales like $(\Omega_bh^2)^
{1/3}\,h^{-3}$.
\par
The results have been numerically computed and are plotted Fig. 5 and Fig. 6, 
integrated over the redshifts $(z>6)$ for different values of 
quasar lifetime ($t_{q}=10^8\,\mbox{yrs}$ and $t_{q}=10^7\,\mbox{yrs}$).
\par
\begin{figure}
\psfig{file=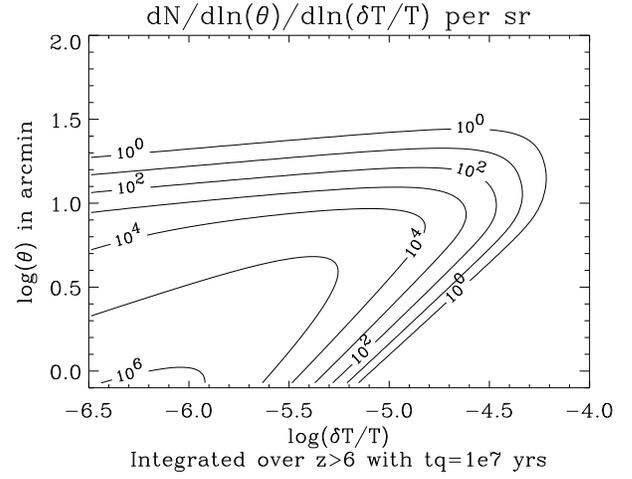,width=\columnwidth}
\caption[ ]{The number of ionized bubbles per unit solid angle, 
logarithmic interval of $|\delta_{T}|$ logarithmic interval of the angular
radius $\theta$ is plotted as a function of the temperature fluctuation 
$\mbox{log}\delta_{T}$ and $\mbox{log}\theta$ in arcmin. This number is 
integrated over the 
redshifts $(z>6)$ and computed for lifetime $t_{q}=10^7\,\mbox{yrs}$ according 
to Eq. (28).}
\end{figure}
\begin{figure}
\psfig{file=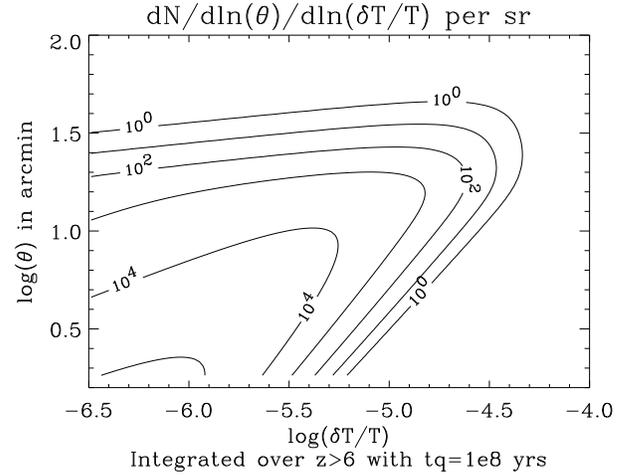,width=\columnwidth}
\caption[ ]{The same quantity is plotted for a quasar lifetime $t_{q}=
10^8\,\mbox{yrs}$. The bubbles are greater but less numerous.}
\end{figure}
\par
Next, we also compute the covering surface parameter $C_{S}$ which is
 the fraction of 
the sky covered by ionized bubbles inducing a distortion of amplitude
 $\delta_{T}$ and angular size $\theta$ at redshift $z$.  This 
is done by multiplying the previous counts by the solid angle occupied by
 the bubbles; we obtain:
\[ C_S=2\,\times\,6\left(\frac{c}{H_0}\right)^{3}\frac{t_0}{t_q}f(z)\,
\times \]
\[ \int\int
\pi\theta^{2}\phi(l_{Q})g(v_{r},z)\,dl_{Q}dv_{r}\,{\bf \delta}[l_{Q}-l_{Q}
(\theta,z)]\, \]
\bequ
{\bf \delta}[\delta_{T}-\delta_{T}(v_{r},l_{Q},z)]
\int_{z}^{+\infty}l(z_{off})\,dz_{off}.
\fequ
Factor 2 takes into account the sources with both negative and positive 
velocities, and $
C_S\propto(\Omega_bh^2)^{-2/3}\,h^{-2}$ for a flat universe.
\par
The results plotted Fig. 7
and Fig. 8 are also integrated over $z$ and given for the two 
previous values of the quasar lifetime $t_{q}$.
\par
\begin{figure}
\psfig{file=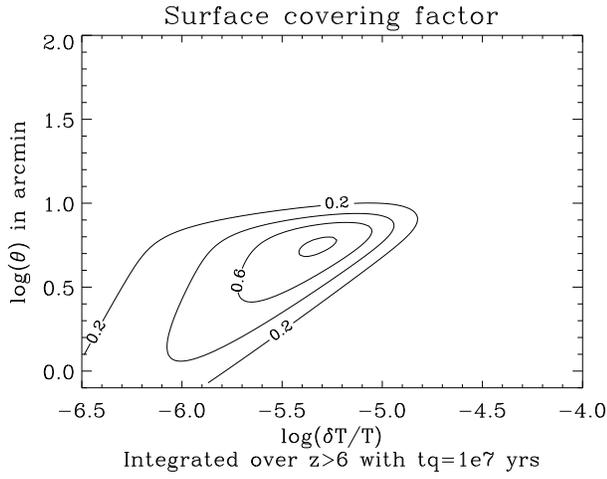,width=\columnwidth}
\caption[ ]{The surface covering factor of the ionized bubbles per
logarithmic interval of $|\delta_{T}|$ and logarithmic interval of the angular
radius $\theta$ is plotted as a function of the temperature fluctuation 
($\mbox{log}\delta_{T}$) and $\mbox{log}\theta$ in arcmin. It is integrated over the 
redshifts $(z>6)$ and computed for lifetime $t_{q}=10^7\,\mbox{yrs}$ according 
to Eq. (29)}
\end{figure}
\begin{figure}
\psfig{file=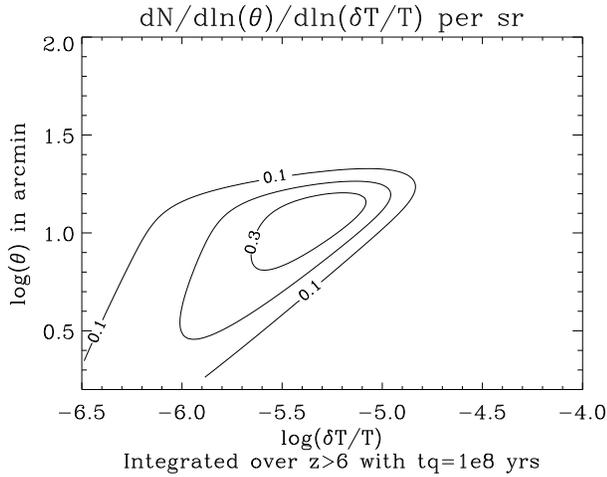,width=\columnwidth}
\caption[ ]{The surface covering factor is plotted for quasar lifetime $t_{q}=
10^8\,\mbox{yrs}$}
\end{figure}
\par
\subsection{Anisotropies during fully ionized period}
When the IGM is totally ionized at $z<z_{ion}=5.7$, we evaluate the rms temperature 
fluctuation. The optical depth is almost constant in this 
redshift range $(z\,<\,z_{ion})$ whereas the velocities are variable. 
Therefore, 
we compute the average of the temperature fluctuation along a line of sight
\bequ
\delta_{T}=\sigma_T\int_{0}^{t_0}n_e v_r\,dt.
\fequ
The fluctuation is due to the peculiar velocity variation which is 
expressed by the velocity dispersion. In the linear perturbation
theory, it is shown that the velocities depend upon the scales of the 
density perturbation that are associated with them and it is also shown that
in a CDM model the coherence length is short. Other cosmological models could 
give a stronger effect.
\par
Using the velocity dispersion of the distribution, we compute the rms
temperature fluctuation $(\delta_T)_{rms}$ when the IGM is fully ionized, 
and find
\bequ
(\delta_T)_{rms}\approx 1.5\,10^{-6}\left (\frac{\Omega_b h^2}{0.015}\right)
\left(\frac{h}{0.5}\right)^{-1}.
\fequ
We note that the effect due to the velocity variation is one order 
of magnitude lower than the rms fluctuation detected by COBE (Mather et al.
1994) and it is at the limit sensitivity of the modern detectors,
the average effect of velocity field in the fully ionized universe is 
therefore weaker than the effect for a partially ionized universe. 
\section{ Discussion}
We discuss here the effects of ionization of the IGM by 
early formed quasars on the cosmic microwave background as calculated in
previous sections.
\par
We have shown the thermal SZ effect was negligible compared to the 
kinetic one. The amplitude of the largest temperature fluctuations 
induced by the latter
mechanism are comparable to the the primordial fluctuations at
scales comparable with the scales of the Doppler peaks ($10^{'}$ to $1^
{\circ}$). 
These secondary fluctuations have no spectral signature
to distinguish them from the primordial ones.
Hence, future observations and
detections of the CMB at small scales with sensitivity about $10^{-6}$ 
will have to take into account the possible presence of such objects 
in the sky. 
\par
When the porosity parameter 
is lower than 1 for $z\,>\,z_{ion}$, we have computed the number of ionized 
bubbles per unit of solid angle, with respect to their angular radius and 
to $\delta_T$ the temperature fluctuation they induce integrating over the
redshifts.
\par
We plot (Fig. 6) the number of kinetic effect
sources integrated over redshifts ($z\,>\,6$) per unit solid angle, first
for lifetime $t_{q}= 10^{8}\,\mbox{yrs}$. We find that there are approximately 
$10^{4 }$ objects per unit solid angle
with associated temperature fluctuations some $10^{-6 }$ and angular
radius 3 to 10 arcmin randomly distributed over the sky and about 10 ionized 
bubbles per unit solid angle with temperature fluctuations  
with about $3\,10^{-5 }$ and radius about 12 to 30 arcmin. With the
same set of parameters, we plot (Fig. 8) the surface covering factor 
integrated over
redshifts versus the amplitude of the anisotropies and their angular radius.
\par
The counts depend significantly on the quasar lifetime value, in fact, it was
shown that decreasing the lifetime $t_{q}$ decreases the size of the bubbles,
but on the other hand, it increases the rate of formation of quasars and thus
gives more numerous bubbles. Figures 5 and 7 show respectively the number
of bubbles per unit solid angle and covering surface versus $\delta_T$ and 
the angular
radius for $t_{q} = 10^{7}\,\mbox{yrs}$. In this configuration, we find that 
there
are 100 objects per steradian with $\delta_T$ about $10^{-5}$ and radius 
5 to 12 arcmin.
\par
Therefore, early quasars will induce distortions with different statistical
properties than the primordial ones and produce rare fluctuations with
amplitude of up to $10^{-5}$.
Although the rms value of the fluctuations induced by the inhomogeneous
reionization is smaller, by an order of magnitude, than the rms of 
the fluctuations detected by COBE (COsmic Background Explorer) 
(Smoot et al. 1992; Mather et al. 1994), the largest ones at small angular 
scales could be dominated by the effect discussed here. Hence high accuracy
measurements of the power spectrum of the CMB anisotropies such 
as those considered for future space experiments (better than 1\%)
and search for deviations to Gaussian statistics must take into account this 
kind of sources.
We find that most of the anisotropies with high level $\Delta T/T$ and small
angular scales are given by quasars at high redshifts.
\par
Observations from the first flight of the Medium Scale Anisotropy 
Measurement (MSAM) (Cheng et al. 1994) with a 30' beam instrument, 
showed the presence of two unresolved sources from a survey of $7.6\,10^{-3}
\,\mbox{sr}$ 
with $\delta_{T}\approx3\,10^{-5}$ consistent spectrally with CMB anisotropies 
; this implies about 200 of such sources per steradian. 
\par 
The temperature fluctuations produced by the effect 
that we are studying -- ionization by early quasars -- are indistinguishable 
from primordial anisotropies. Furthermore, the computations done for our
model with short quasar lifetime
$t_{q} = 10^{7}\,\mbox{yrs}$ predicts about 100 events 
per unit solid angle with angular diameter about 20' and 
$\delta_{T}\approx3\,10^{-5 }$, consistent with the results of the 
MSAM observations. In this context, the statistics of such events on the sky 
will be an important criterion to attribute an origin to them.
\par
Figure 7 -- representing the surface covering factor -- shows that the ionized 
bubbles around quasars are close to cover the whole 
sky although they do not fill the space until $z\,<\,z_{ion}=5.7$.
Therefore, we can no longer consider the induced temperature fluctuations as 
individual sources. On the contrary, it is necessary to get statistical 
informations about the distribution of the anisotropies such as the standard 
deviation, mean value, kurtosis... Hence we have simulated maps
of the anisotropies induced by the quasars.
\par
The maps were drawn under simple geometrical assumptions (spherical bubbles), 
using the counts of ionized bubbles associated with quasars characterized by
an angular radius and a temperature fluctuation given in Eq. (28) and 
using the fact that there is a simple relation between the angular radius 
of the bubble
and the luminosity of the quasar that induced it, Eq. (12). We put a limit 
on the number of bubbles; for numbers of bubbles greater 
than this limit, a gaussian drawing is done otherwise we make a poissonnian 
drawing. The positions of the bubbles over the map are random, this is a
rather good assumption since the clustering of quasars is still uncertain.
\par
We check that the standard deviation, mean, maximum and minimum values are found
consistent with the bubbles counts. But as could be expected, the distribution 
of the temperature fluctuations induced by quasars
is far from being gaussian. Indeed, for a 
typical value of the quasar lifetime $t_q=10^7\,\mbox{yrs}$ we find that the 
kurtosis
is equal to 1.4. This is an important result that might be taken into account
if one wants to test the gaussiannity of the primordial anisotropies of the
CMB, to prove whether or not the primordial fluctuations are adiabatic (thus 
due to quantum
fluctuations of the vacuum energy) for which the statistic is gaussian or to 
topological defaults (strings, domain walls, ...). 
In fact, the presence of early ionizing sources (quasars here) induces
additional non gaussian anisotropies with no spectral signature. 
\par
Cosmological models show that there is an angular cut off in the spectrum of 
the primordial anisotropies. The bubbles contribution to the observed CMB at 
small angular scales could be important or may be dominant, 
in particular at scales where the primordial fluctuations are 
"washed out" because of the damping associated with the width of the 
the last scattering surface. This effect is suggested Fig. 9, where we plot 
both the power spectra of the primordial temperature fluctuations and the
bubbles in terms of the spherical harmonics coefficients (see for example 
White, Scott \& Silk 1994); taking $t_q=10^7\,
\mbox{yrs}$.
\begin{figure}
\psfig{file=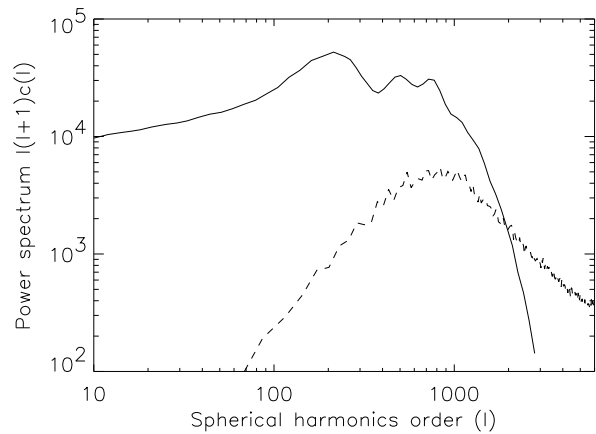,width=\columnwidth}
\caption[ ]{Power of the anisotropies induced by the bubbles 
 and the primordial fluctuations in the spherical harmonics
versus the multipoles. {\it Solid curve} for the primary CMB and {\it dashed 
curve} for the fluctuations induced by the quasars we take $t_{q}=10^7\,
\mbox{yrs}$}
\end{figure}
Our results show that the secondary anisotropies of the CMB, produced 
by the kinetic SZ effect in ionized bubbles around quasars, must be taken 
into account. 
\par

In the future, observations along the lines of sight of known quasars 
at high redshift showing the Gunn-Peterson effect, could be an interesting test
to detect this kinetic effect induced by the ionized bubbles.
The following calculations are possible only for the bubbles associated with 
the strongest temperature fluctuations for which primordial anisotropies do not 
perturb too much the observations. In that case, we could be able to derive the 
bubble peculiar velocity.
\par
In Sect. 2.6, We have derived a simple relation between the temperature 
fluctuation associated with an ionized bubble, the radial peculiar velocity 
and the luminosity of the quasar that has produced it Eq. (18). 
On the other hand, Eq. (12) gives the the expression 
of the angular radius of the the bubble as a function of the quasar luminosity.
Therefore, if we observe a bubble inducing a temperature fluctuation, we 
directly derive the
luminosity from the angular size and knowing this we have a direct measure 
of the peculiar radial velocity. The latter is given by the following
expression 
\bequ
\frac{v_r}{c}=\frac{3}{2}\frac{\delta_T}{\theta}\frac{\theta_c}{\tau_c}
(1+z)^{-2}\,[1-(1+z)^{1/2}]^{-1},
\fequ
where $\theta_c$ and $\tau_c$ are defined respectively in Sect. 2.3. and Sect.
2.4.
\par
In the hypothesis of an ionization due to individual sources of radiation, 
the whole formalism we develop 
remains valid whatever the kind of sources, early galaxies, first generation 
of stars, ... The effect of the presence of sources of luminosity of the same
order as the quasars on the CMB will be the same as the one described for 
quasars. 
\begin{acknowledgements}
The authors thank M. Lachi\`eze--Rey and F. R. Bouchet for helpful discussions,
and an anonymous referee for helpful comments.
\end{acknowledgements}
\end{document}